\documentclass[conference]{IEEEtran}
\IEEEoverridecommandlockouts

\usepackage{cite}
\usepackage{amsmath,amssymb,amsfonts}
\usepackage{algorithm}
\usepackage{algpseudocode}
\usepackage{graphicx}
\usepackage{textcomp}
\usepackage{xcolor}
\usepackage{bm}
\usepackage{booktabs}
\usepackage{url}
\usepackage[hidelinks]{hyperref}

\newcommand{\HH}{\mathbf{H}}
\newcommand{\vv}{\mathbf{v}}
\newcommand{\nn}{\mathbf{n}}
\newcommand{\gam}{\boldsymbol{\gamma}}
\newcommand{\Kg}{\mathbf{K}_\gamma}
\newcommand{\Kn}{\mathbf{K}_n}
\newcommand{\Ke}{\mathbf{K}_\epsilon}
\newcommand{\WMMSE}{\mathbf{W}_{\mathrm{MMSE}}}
\newcommand{\II}{\mathbf{I}}
\newcommand{\herm}{^{\mathsf{H}}}
\newcommand{\trans}{^{\mathsf{T}}}
\newcommand{\tr}{\mathrm{Tr}}
\newcommand{\E}{\mathbb{E}}

\newcommand{\C}{\mathbb{C}}
\newcommand{\Order}{\mathcal{O}}

\def\BibTeX{{\rm B\kern-.05em{\sc i\kern-.025em b}\kern-.08em
    T\kern-.1667em\lower.7ex\hbox{E}\kern-.125emX}}

\begin{document}

\title{Iterative Reduced-Rank MMSE Estimation of Sparse Range Profiles from Non-Contiguous Radar Transmission Spectra}

\author{
\IEEEauthorblockN{
Gordon Ariho\IEEEauthorrefmark{1}, 
Hara Madhav Talasila\IEEEauthorrefmark{1}, 
James M. Stiles\IEEEauthorrefmark{2}, and 
Peng Seng Tan\IEEEauthorrefmark{1}
}

\IEEEauthorblockA{\IEEEauthorrefmark{1}Independent Researcher (Research conducted at the University of Kansas, Lawrence, KS, USA)}
\IEEEauthorblockA{\IEEEauthorrefmark{2}Radar Systems and Remote Sensing Laboratory (RSL), University of Kansas, Lawrence, USA}
\IEEEauthorblockA{Corresponding Author: gordon.ariho@gmail.com}
}

\maketitle

\begin{abstract}
Ongoing demand for radio spectrum by commercial wireless services has steadily increased pressure on the frequency bands traditionally reserved for radar. 
This paper addresses the joint problem of designing non-contiguous radar transmission spectra and estimating the range profile from the resulting reduced measurement set.
Transmission spectra are constructed using a Marginal Fisher Information (MFI) criterion that removes blocks of frequencies contributing least to estimation accuracy. 
To process the underdetermined signals acquired from the resulting sparse measurement vector, an iterative Reduced-Rank Minimum Mean-Square Error (RRMMSE) estimator is proposed. 
The estimator starts with a single-target hypothesis and grows the active target subspace one range bin at a time, updating the a~priori target covariance matrix in each iteration using both the largest estimated reflection coefficient and its posterior error variance.
This avoids inversion of the full $M{\times}M$ covariance matrix that would be required by a one-step MMSE and concentrates the rank of the estimator on the support of significant scatterers. 
The Bayesian Cram\'er--Rao Lower Bound (CRLB) on the per-bin reflection coefficient is derived for the non-contiguous spectrum measurement model, and the computational complexity of the proposed estimator is shown to scale as $\Order(G^2 M K^2)$, where $G$ is the number of detectable scatterers, $M$ is the number of range bins, and $K$ is the number of preserved spectral samples. 
Simulations using $50\%$ and $75\%$ spectrally occupied MFI-designed spectra confirm that the algorithm recovers sparse range profiles with Mean-Square Error (MSE) close to the fully filled baseline when the number of significant scatterers is not larger than the rank of the sparse sensing matrix.
\end{abstract}


\section{Introduction}\label{sec:intro}

\IEEEPARstart{R}{adar} systems have historically benefited from generous, often exclusive, frequency allocations within the UHF or SHF bands, that include IEEE L, S, and X bands. 
With the rapid proliferation of commercial wireless devices and the growing demand for bandwidth from 5G and future 6G networks, this privileged status can no longer be sustained indisputably~\cite{Griffiths_2015,Haykin_2006}.
Regulatory bodies have moved towards Dynamic Spectrum Access (DSA), forcing radar systems either to vacate portions of their historical bands or to share them with primary or secondary commercial users~\cite{FCC_3p5GHz_2015,Cohen_2018}. 
On the complementary side of dynamic spectrum access, the secondary users that occupy the relinquished bands must first sense the wideband spectrum to identify the vacated spectrum holes, a task recently addressed with collaborative and federated machine-learning frameworks for networked unmanned aerial vehicles operating within unmanned aircraft system traffic management ecosystems~\cite{Chintareddy_GLOBECOM_2023,Chintareddy_TMLCN_2025}. 

The pressure is particularly evident for scientific airborne radar systems whose science return scales with instantaneous bandwidth. 
Ultra-wideband UHF ice-penetrating radars developed for the radar remote sensing of the cryosphere~\cite{Morales_IGARSS_2024_Thwaites} and multi-channel ultra-wideband airborne radars used for swath mapping of snow layers~\cite{Morales_IGARSS_2024_Multichannel} occupy several hundred megahertz of contested frequency. 
Comparable bandwidth demands arise in high-altitude snow-thickness sensing with ultra-wideband microwave radar~\cite{Talasila_IGARSS_2023}.
For such instruments, the option to vacate or share portions of the band without sacrificing range resolution or amplitude fidelity is of immediate practical interest. 

Within radar array processing, \emph{spectral efficiency} can be understood as the optimal use of bandwidth to achieve estimation performance comparable to the fully filled bandwidth case, while freeing portions of the spectrum that can be re-allocated to other wireless emitters. 
Achieving spectral efficiency requires both (i)~a transmission spectrum that is informationally optimal subject to a sparsity constraint, and (ii)~a receive processor that can recover the range profile from the resulting reduced measurement set without unacceptable loss of accuracy.
In quantitative remote-sensing applications, the recovered range profile is needed in addition to a single surface scatterer measurement. The radiometric calibration of radar depth sounder data products~\cite{Talasila_RadarConf_2024} and the downstream retrieval of geophysical parameters depend on the amplitude of every detected scatterer, raising the standard for what is considered an acceptable loss of accuracy.

\subsection{Related Work}

Spectrum-sharing radar literature can be broadly classified into three categories.
\emph{Waveform-design} approaches optimize the transmitted waveform under spectral compatibility constraints, with notches placed at frequencies allocated to other
users~\cite{Aubry_2014,Blunt_Mokole_2016}. 
\emph{Joint radar communication} and integrated sensing and communication approaches design dual-function waveforms that simultaneously serve sensing and data transmission~\cite{Liu_2020,Liu_2022,Mishra_2019}.
\emph{Compressive sensing} radar exploits sparsity of the range or range-Doppler profile to recover targets from sub-Nyquist measurement sets~\cite{Candes_2006,Ender_2010,Cohen_2018,Baraniuk_2007,Quan_2024}.
Recent work has also explored cognitive radar resource management with feedback between the receiver and the waveform generator~\cite{Greco_2018,Charlish_Hoffmann_2020}. 
From the perspective of the secondary network that reuses the spectrum vacated by such radars, data-driven wideband spectrum sensing combined with reinforcement-learning-based scheduling has been proposed to detect and allocate the available spectrum holes among networked unmanned aerial vehicles~\cite{Chintareddy_GLOBECOM_2023,Chintareddy_TMLCN_2025}. 
Bayesian covariance-based estimation approaches related to sparse inverse radar problems have also been explored in multipass interferometric radar processing and radar tomography applications~\cite{Ariho_PhD_2023,Ariho_PAST2022,Ariho_IGARSS2020}, including the joint recovery of ice-sheet vertical velocity, englacial layer geometry, and catchment-scale englacial deformation from repeated synthetic aperture radar passes~\cite{Ariho_PAST2022,Holschuh_AGU_2022}.
Two recent surveys that are related include an ISAC-oriented review of dual-functional networks for the 6G era~\cite{Liu_2022} and a survey on sparse-recovery techniques in modern radar signal processing~\cite{Quan_2024}.

The work presented here belongs to the \emph{information-theoretic} line introduced in~\cite{Tan_2016,Stiles_2009}, in which the spectral support itself is selected to maximize estimation information about the range profile, rather than being treated as a fixed constraint. 
Compared with greedy sparsity-recovery algorithms that operate on a fixed measurement set, the Marginal Fisher Information (MFI) driven design treats the spectral support as a degree of freedom, and the resulting sparse arrays exhibit coarrays with low redundancy and few holes~\cite{Tan_2016,Holm_1989,Moffet_1968}.

Beyond the spectrum-sharing literature itself, the receive-side estimation problem treated here is increasingly coupled to community-scale scientific products. 
Open infrastructure for the standardized referencing, processing, and access of radar echograms~\cite{Paden_OPR_2022,Paden_AGU_2021} now exposes the per-bin reflection coefficient to a wide range of scientific communities, including deep-neural-network layer trackers trained on NASA Operation IceBridge data~\cite{Oluwanisola_RadarConf_2023}, AI-ready snow-radar datasets curated for climate-change monitoring~\cite{Oluwanisola_SRED_2025}, and paleoclimate age-scale reconstructions from new ice cores~\cite{Rowell_Climate_2023}. 
The same echogram products also underpin glaciological interpretation at the catchment scale, as in the three-dimensional swath-radar imaging of subglacial topography used to infer the tectonic controls on ice-sheet nucleation in West Antarctica~\cite{Hoffman_NatGeo_2023}.
Estimation errors introduced upstream by a sparse transmission spectrum therefore propagate into derived geophysical quantities, which reinforces the treatment of the posterior error covariance studied in this paper.

\subsection{Contributions}

This paper expands on the spectrum design framework of \cite{Tan_2016,Stiles_2009} and proposes a complementary receiver-side processor. 
The specific contributions are summarized as follows.
\begin{enumerate}
\item The Bayesian Fisher Information Matrix (FIM) and the associated Cram\'er--Rao Lower Bound (CRLB) on the complex reflection coefficient at each range bin are derived for the non-contiguous spectrum measurement model. 
This work defines the link between the trace of the Bayesian CRLB and the MFI criterion of Tan~\cite{Tan_2016}. 

\item An iterative reduced-rank MMSE estimator is proposed in which the a~priori target covariance matrix is updated one range bin at a time using both the estimated reflection coefficient and the posterior error variance. 
The rank of the estimator is grown only along the support of significant scatterers, which avoids the $\Order(M^3)$ inversion required by the full one-step MMSE.

\item The computational complexity of the proposed estimator is analyzed and compared with that of Orthogonal Matching Pursuit (OMP)~\cite{Tropp_Gilbert_2007}, the CLEAN algorithm~\cite{Bose_2012}, and Sparse Bayesian Learning (SBL)~\cite{Wipf_Rao_2004,Tipping_2001}. 
The proposed estimator is shown to provide a clear posterior error covariance that serves both as a stopping criterion and as the input to the Bayesian update of the prior, a property not shared with the listed alternatives.

\item The algorithm is evaluated on MFI-designed sparse spectra at $50\%$ and $75\%$ spectral occupancy across a range of target densities. 
The achieved MSE is shown to approach that of the fully filled spectrum whenever the number of significant scatterers is at most equal to the rank of the sparse sensing matrix.
\end{enumerate}

The remainder of the paper is organized as follows. 
Section~\ref{sec:formulation} states the linear measurement model. 
Section~\ref{sec:spectrum} reviews the MFI criterion for designing non-contiguous spectra and derives the Bayesian CRLB. 
Section~\ref{sec:algorithm} presents the iterative RRMMSE algorithm, analyzes its computational complexity, and relates it to other greedy sparse-recovery algorithms. 
Section~\ref{sec:results} reports simulation results, 
and Section~\ref{sec:conclusion} concludes the paper.

\section{Problem Formulation}\label{sec:formulation}
\subsection{Spectral Model}

Consider a pulsed radar that generally occupies a bandwidth $B$ centered at a frequency $f_c$. The transmit spectrum is modeled as a sum of Dirac delta functions located at $N$ uniformly spaced frequencies separated by $\Delta\omega$:
\begin{equation}
S(\omega) \;=\; \sum_{n=0}^{N-1} s_n\, \delta(\omega - n\Delta\omega),
\label{eq:spectrum_model}
\end{equation}
where $s_n$ denotes the complex weight of the $n$-th spectral line. 
In a non-contiguous spectrum, a subset of the weights $\{s_n\}$ is set to zero to vacate frequencies allocated to other users. 
The number of the remaining (nonzero) lines is denoted $K$. 
Each spectral line can be viewed as an independent narrow-band coherent transmit/receive channel, so the full transmit spectrum is equivalent to a uniform linear array (ULA) of $N$ elements, and the non-contiguous spectrum is equivalent to a sparse, non-uniform linear array of $K$ elements, where $K<N$. 

\subsection{Linear Measurement Model}
The discrete range axis has $M$ resolution cells with known centers $\tau_1, \tau_2, \dots, \tau_M$, and let $\gamma_m \in \C$ denote the complex reflection coefficient associated with the $m$-th range cell. 
The discrete radar measurement vector, $\vv\in\C^{K\times 1}$, obtained at the preserved
spectral lines is then

\begin{equation}
\vv \;=\; \HH\,\gam + \nn,
\label{eq:linear_model}
\end{equation}

where:
\begin{itemize}
\item $\gam = [\gamma_1, \gamma_2, \dots, \gamma_M]\trans \in \C^{M\times 1}$ is the range-profile vector.
\item $\HH\in\C^{K\times M}$ is the sensing matrix whose $(k,m)$ element is

\begin{equation}
[\HH]_{k,m} \;=\; \exp\!\left(j\,\omega_{k}\,\tau_m\right),
\label{eq:Hkm}
\end{equation}

where $\{\omega_k\}_{k=1}^{K}$ are the preserved angular frequencies selected from the $N$ lines. 

\item $\nn\in\C^{K\times 1}$ is the additive measurement noise, modeled as zero-mean circularly symmetric complex Gaussian with covariance matrix $\Kn = \sigma_n^2\,\II_K$, where $\sigma_n^2$ is variance or noise power and $\II_K$ is $K\times K$ identity matrix.
\end{itemize}

The columns of $\HH$ are normalized to unit Euclidean norm so that the diagonal of $\HH\herm\HH$ is unity for all spectral occupancy levels, isolating the effect of sparsity from gross signal-power scaling.

\subsection{The Sparsity Problem}
When the transmit spectrum is non-contiguous with $K<M_{\mathrm{nyq}}$, where $M_{\mathrm{nyq}} \leq M$ is the number of Nyquist frequency samples in $B$, the Gram matrix $\HH\herm\HH$ is no longer a diagonal matrix. 
Off-diagonal entries grow in magnitude as the preserved frequencies are removed, raising the integrated sidelobe level (ISL) of the matched-filter response and increasing the estimation error variance of $\gam$. 
Furthermore, when $K<M$, the problem is underdetermined and the matched filter (i.e., back-projection $\HH\herm\vv$) suffers from large leakage between range cells. 

The range profile $\gam$, however, is typically sparse and most elements correspond to resolution cells with no detectable target. 
This sparsity assumption is well-supported in many practical settings; for example, airborne ice-sounding waveforms typically yield only a few range bins of strong returns per pulse that include air--ice interface, a limited set of internal ice layers, and the ice--bedrock interface~\cite{Talasila_RADAR_2025}, and the suppression of off-nadir clutter that masks these returns has been treated through multipass tomographic processing of radar depth sounder data~\cite{Ariho_IGARSS2020}.
Let $G$ denote the number of resolution cells with significant reflectivity. When $G \leq K$, the problem of recovering $\gam$ from $\vv$ is informationally feasible provided that the support of $\gam$ can be identified.  
The algorithm proposed in Section~\ref{sec:algorithm} addresses exactly this support identification problem by iteratively constructing a reduced-rank subspace of $\HH$ where significant scatterers are most likely to reside.

\section{Spectrum Design and Bayesian CRLB}\label{sec:spectrum}

The MMSE and CRLB results below are classical Bayesian estimation theory~\cite{Kay_1993}, restated here in the notation of the non-contiguous spectrum model~\eqref{eq:linear_model}. 
The MFI criterion is adopted from~\cite{Tan_2016,Stiles_2009}. 
The derivation of the Bayesian CRLB for the complex reflection coefficients under the spectrum model~\eqref{eq:Hkm} and the connection between that bound and the MFI criterion are stated as new results in Section~\ref{subsec:crlb} and Section~\ref{subsec:mfi_link}.

\subsection{Posterior Statistics and the MMSE Estimator}

We assign a zero-mean circular complex Gaussian prior on $\gam$ with covariance $\Kg = \E\{\gam\gam\herm\} = \sigma_\gamma^2\,\II_M$.
Conditioned on $\gam$, the measurement $\vv$ is complex Gaussian with mean $\HH\gam$ and covariance $\Kn$. 
The classical MMSE estimator of $\gam$ given $\vv$ is~\cite[Ch.~11]{Kay_1993}:
\begin{equation}
\WMMSE = \Kg\HH\herm
\left(\HH\Kg\HH\herm + \Kn\right)^{-1},
\label{eq:wmmse}
\end{equation}

\begin{equation}
\hat\gam_{\mathrm{MMSE}} \;=\; \WMMSE\,\vv.
\label{eq:gamma_mmse}
\end{equation}

The posterior error covariance is
\begin{align}
\Ke
&= \Kg - \Kg\HH\herm
\left(\HH\Kg\HH\herm + \Kn\right)^{-1}\HH\Kg
\label{eq:Ke_form1} \\
&= \left(\HH\herm\Kn^{-1}\HH + \Kg^{-1}\right)^{-1}.
\label{eq:Ke_form2}
\end{align}

Equation~\eqref{eq:Ke_form2} follows from~\eqref{eq:Ke_form1} by the matrix inversion lemma. For the white-noise, uniform-prior case considered throughout this paper this reduces to
\begin{equation}
\Ke \;=\; \left(\sigma_n^{-2}\,\HH\herm\HH + \sigma_\gamma^{-2}\,\II_M\right)^{-1}.
\label{eq:Ke_white}
\end{equation}

\subsection{Bayesian Fisher Information and CRLB}
\label{subsec:crlb}

The radar target parameter of interest in this work is the complex reflection coefficient $\gamma_m$ at each candidate range bin~$m$. 
The delay $\tau_m$ is the center of the $m$-th range bin and is treated as known a~priori; estimation of $\tau_m$ itself would require a denser range-bin grid and is outside the scope of this paper.

For the measurement model in~\eqref{eq:linear_model} with circular complex Gaussian noise, the deterministic complex Fisher Information Matrix (FIM) of $\vv$ with respect to $\gam$ is~\cite[Ch.~15]{Kay_1993}
\begin{equation}
\mathbf{J}(\gam) \;=\; \HH\herm\,\Kn^{-1}\,\HH.
\label{eq:fim_det}
\end{equation}

When $\gam$ is assigned the Gaussian prior $\Kg$, the \emph{Bayesian} FIM is
\begin{equation}
\mathbf{J}_B(\gam)
\;=\; \HH\herm\Kn^{-1}\HH \;+\; \Kg^{-1}.
\label{eq:fim_bayes}
\end{equation}

The Bayesian CRLB on the error covariance of any estimator of $\gam$ given $\vv$ is
\begin{equation}
\Ke \;\succeq\; \mathbf{J}_B^{-1}
\;=\; \left(\HH\herm\Kn^{-1}\HH + \Kg^{-1}\right)^{-1},
\label{eq:bcrlb}
\end{equation}

with the inequality interpreted in the positive semi-definite sense. 
The bound on the mean square error of the $m$-th reflection coefficient is the $m$-th diagonal element of the inverse,
\begin{equation}
\mathrm{CRLB}(\gamma_m)
\;=\; \left[\left(\HH\herm\Kn^{-1}\HH + \Kg^{-1}\right)^{-1}\right]_{m,m}.
\label{eq:bcrlb_diag}
\end{equation}

Comparing~\eqref{eq:bcrlb} with~\eqref{eq:Ke_form2} shows that the MMSE estimator~\eqref{eq:wmmse} attains the Bayesian CRLB whenever the prior $\Kg$ matches the true second-order statistics of $\gam$. 
The role of the iterative algorithm in Section~\ref{sec:algorithm} is precisely to refine the prior $\Kg$ so that this match is approached, even when only a coarse initial prior is available.

\subsection{Marginal Fisher Information and the Link to CRLB}
\label{subsec:mfi_link}

The MFI metric measures the change in information gained when a single new block of spectral lines is added to the existing sparse spectrum. 
Let $\mathbf{J}_{B,K}$ denote the Bayesian FIM in~\eqref{eq:fim_bayes} when $K$ spectral lines are preserved, and $\mathbf{J}_{B,K-1}$ the corresponding FIM with one block removed. 
Following~\cite{Tan_2016,Stiles_2009}, the MFI matrix obtained from the $K$-th block is the nonnegative-definite matrix
\begin{equation}
\Delta \mathbf{I}(K)
\;=\; \mathbf{J}_{B,K-1}^{-1} - \mathbf{J}_{B,K}^{-1},
\label{eq:mfi_matrix}
\end{equation}

and the scalar MFI used as the design criterion is
\begin{equation}
\mathrm{MFI}(K) \;=\; \frac{1}{M}\,\tr\!\left(\Delta\mathbf{I}(K)\right).
\label{eq:mfi_scalar}
\end{equation}

Since $\mathbf{J}_{B,K}^{-1}$ is the Bayesian CRLB matrix at sparsity level $K$, equation~\eqref{eq:mfi_scalar} can be written equivalently as
\begin{equation}
\mathrm{MFI}(K) \;=\; \frac{1}{M}\Big[\tr(\Ke^{(K-1)})
- \tr(\Ke^{(K)})\Big],
\label{eq:mfi_in_Ke}
\end{equation}

where $\Ke^{(K)}$ is the posterior error covariance obtained when $K$ spectral lines are preserved. The MFI criterion is therefore the marginal reduction in the trace of the Bayesian CRLB matrix per added block. 
Selecting frequency blocks to maximize MFI is equivalent to selecting blocks that most reduce the trace of the Bayesian CRLB, which is the connection between the spectrum-design problem of~\cite{Tan_2016} and the receive-side estimation problem treated here.

\begin{figure}[!t]
\centerline{\includegraphics[width=3.4in]{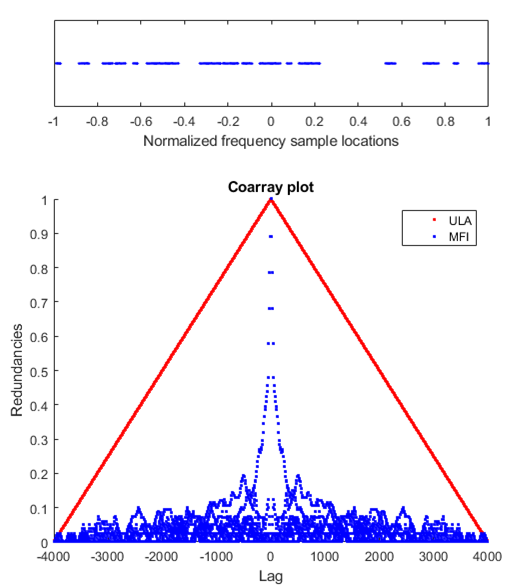}}
\caption{MFI-designed sparse spectrum at $50\%$ spectral occupancy. Top: locations of preserved frequency samples (normalized to the edge of the band). Bottom: coarray $c(l)$, normalized to the zero-lag value, for the full ULA (red squares) and for the MFI-designed sparse spectrum (blue squares). The MFI design achieves coverage at every nonzero lag with low redundancy.}
\label{fig:50_spectrum}
\end{figure}

\begin{figure}[!t]
\centerline{\includegraphics[width=3.4in]{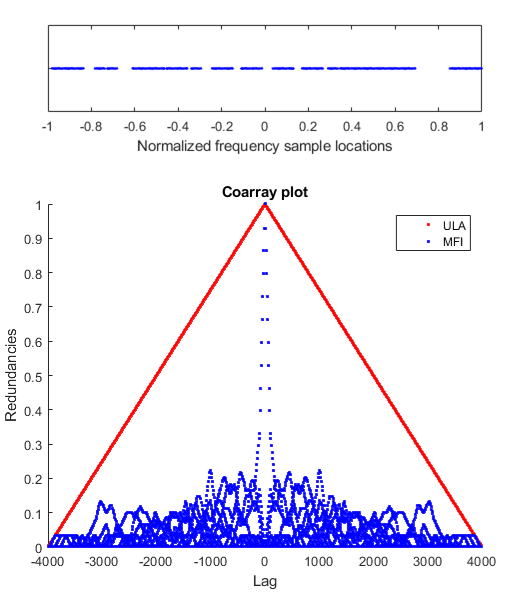}}
\caption{MFI-designed sparse spectrum at $75\%$ spectral occupancy. Top: preserved frequency samples. Bottom: coarray of the full ULA (red squares) and of the MFI-designed sparse spectrum (blue squares). Increased occupancy reduces the gap density in the spectrum and improves the coarray redundancy profile.}
\label{fig:75_spectrum}
\end{figure}

The sparse spectra used in this paper are generated using the block-removal MFI optimization detailed in~\cite{Tan_2016}, with a block size of $1.25\%$ of the filled spectrum. 
The interval between adjacent samples within each block satisfies the Nyquist criterion. 
The co-array 
\begin{equation}
c(l) \;=\; \sum_{n=0}^{N-l-1} w_n\,w_{n+l},
\label{eq:coarray}
\end{equation}

where $w_n\in\{0,1\}$ encodes the presence or absence of each spectral line, is used to verify that the MFI-designed support has low redundancy and few holes~\cite{Holm_1989,Moffet_1968}.
Fig.~\ref{fig:50_spectrum} and Fig.~\ref{fig:75_spectrum} show the MFI-designed spectral supports at $50\%$ and $75\%$ occupancy, respectively, together with their coarrays.

\section{Iterative Reduced-Rank MMSE}\label{sec:algorithm}

A direct application of the one-step MMSE filter~\eqref{eq:wmmse} to a sparse-spectrum measurement $\vv$ requires inversion of the $K\times K$ matrix $\HH\Kg\HH\herm + \Kn$, but with $\Kg = \sigma_\gamma^2\II_M$ this spreads the prior power uniformly across all $M$ range bins, most of which contain no detectable scatterer. 
The resulting estimate $\hat\gam_{\mathrm{MMSE}}$ tends to either miss weak targets or produce many false alarms in the low-sample-support regime $K<M$. 
The proposed iterative scheme replaces the flat prior $\Kg$ by a sequence of progressively refined priors that concentrate target power on bins identified during previous iterations.

\subsection{Algorithm Statement}

Let $\mathcal{S}^{(i)}\subseteq\{1,2,\dots,M\}$ denote the estimated target support after iteration~$i$, with $\mathcal{S}^{(0)}=\varnothing$.
At iteration~$i$, the algorithm maintains a diagonal target covariance matrix $\Kg^{(i)}$ whose $m$-th diagonal entry is
\begin{equation}
[\Kg^{(i)}]_{m,m}
\;=\;
\begin{cases}
|\hat\gamma_m^{(i)}|^2 + [\Ke^{(i)}]_{m,m}, &
m\in\mathcal{S}^{(i)},\\[2pt]
\sigma_\gamma^2,& \text{otherwise},
\end{cases}
\label{eq:Kg_update}
\end{equation}

where $\hat\gamma_m^{(i)}$ is the running estimate of the reflection coefficient at bin~$m$ and $[\Ke^{(i)}]_{m,m}$ is its posterior error variance, both available in closed form from~\eqref{eq:wmmse}--\eqref{eq:Ke_form2}.
\begin{algorithm}[!t]
\caption{Iterative Reduced-Rank MMSE (RRMMSE)}
\label{alg:rrmmse}
\begin{algorithmic}[1]
\Require Measurement $\vv$, sensing matrix $\HH$,
noise covariance $\Kn$, prior variance $\sigma_\gamma^2$,
maximum iterations $G_{\max}$, tolerance $\eta$.
\State Initialize $\Kg^{(0)}\gets\sigma_\gamma^2\II_M$,
$\mathcal{S}^{(0)}\gets\varnothing$, $i\gets 1$.
\Repeat
\State \textbf{Test each candidate bin:}
for every $m\notin\mathcal{S}^{(i-1)}$, form the augmented
index set $\mathcal{I}_m = \mathcal{S}^{(i-1)}\cup\{m\}$ and
evaluate
\[
\hat\gamma_{(m)}
\;=\;
\Kg^{(i-1)}\,
\HH(:,\mathcal{I}_m)\herm
\left(\HH\Kg^{(i-1)}\HH\herm + \Kn\right)^{-1}\vv .
\]
\State \textbf{Pick the strongest candidate:}
$m^\star \gets \arg\max_{m\notin\mathcal{S}^{(i-1)}}
\big|[\hat\gamma_{(m)}]_{\,m}\big|$ and set
$\mathcal{S}^{(i)} \gets \mathcal{S}^{(i-1)} \cup \{m^\star\}$.
\State \textbf{Update posterior covariance}
$\Ke^{(i)}$ via~\eqref{eq:Ke_form1}.
\State \textbf{Update prior} $\Kg^{(i)}$ using~\eqref{eq:Kg_update}.
\State $i\gets i+1$.
\Until{$\tfrac{1}{M}\tr(\Ke^{(i-1)})$ stops decreasing
within tolerance $\eta$, or $i>G_{\max}$.}
\State \Return Support $\mathcal{S}^{(i-1)}$ and estimate
$\hat\gam^{(i-1)}$.
\end{algorithmic}
\end{algorithm}

\begin{figure}[!t]
\centerline{\includegraphics[width=3.0in]{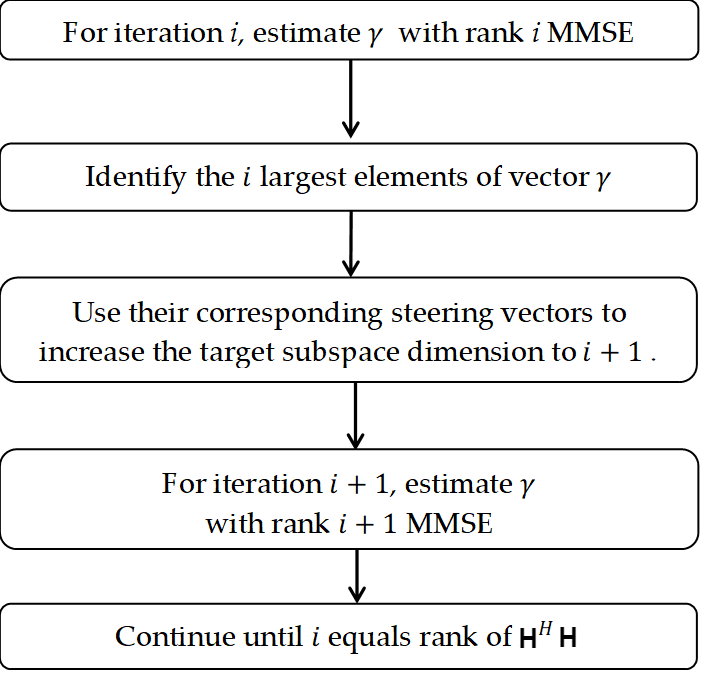}}
\caption{Flow chart of the iterative Reduced-Rank MMSE algorithm.
The rank of the estimator is grown one range bin per iteration along the support identified by the magnitude of the running reflection-coefficient estimate.}
\label{fig:flowchart}
\end{figure}

Algorithm~\ref{alg:rrmmse} summarizes the procedure.
Starting with the rank-one hypothesis, the algorithm at each iteration tests every range bin that is not yet in the support, retains the bin that yields the largest reflection coefficient under the current augmented prior, and updates the prior by absorbing the new bin into the support.
The iterations terminate when the trace of the posterior error covariance, $\tfrac{1}{M}\tr(\Ke^{(i)})$, stops decreasing within a chosen tolerance.
The flow chart of this procedure is reproduced in Fig.~\ref{fig:flowchart}. 

\subsection{Computational Complexity}
\label{subsec:complexity}

Let $G$ denote the final number of iterations executed by Algorithm~\ref{alg:rrmmse}.
By construction, $G\leq K$, since the rank of the sparse sensing matrix bounds the number of scatterers that can be estimated from $\vv$.
At iteration~$i$:
\begin{enumerate}
\item The inverse $\left(\HH\Kg^{(i-1)}\HH\herm
+ \Kn\right)^{-1}$ is formed once. This is a $K\times K$ matrix and the inversion costs $\Order(K^3)$ flops.
It depends on the running $\Kg^{(i-1)}$ but not on the candidate index~$m$.
\item For each of the $M-(i-1)$ candidate bins, the candidate estimate of length~$i$ in step~3 of Algorithm~\ref{alg:rrmmse} costs $\Order(iK^2)$ flops.
The total cost over all candidates at iteration~$i$ is $\Order(MiK^2)$.
\item The prior update~\eqref{eq:Kg_update} costs $\Order(M)$ flops.
\end{enumerate}

The dominant per-iteration cost is therefore $\Order(K^3 + MiK^2)$.
Summing $i$ from $1$ to $G$ gives the overall cost
\begin{equation}
T_{\mathrm{RRMMSE}}
\;=\;
\Order\!\left(G\,K^3 + G^2 M K^2\right).
\label{eq:complexity}
\end{equation}
Since $G\leq K \leq M$ in the regime of interest, the second term
dominates and the complexity reduces to
\begin{equation}
T_{\mathrm{RRMMSE}} \;=\;
\Order(G^2 M K^2).
\label{eq:complexity_dominant}
\end{equation}

For comparison, a one-step full-rank MMSE applied to the same measurement vector requires inversion of the $M\times M$ covariance matrix in~\eqref{eq:Ke_form2}, at a cost of $\Order(M^3)$.
In the sparse-target regime considered here, where the number of significant scatterers $G$ is much smaller than $M$, the proposed estimator is therefore substantially cheaper.
OMP has complexity $\Order(GMK + G^4)$~\cite{Tropp_Gilbert_2007}; the proposed estimator does more work per iteration than OMP because it carries the full prior $\Kg^{(i)}$, but in return it provides the closed-form posterior covariance $\Ke^{(i)}$ used for both the prior update~\eqref{eq:Kg_update} and the stopping criterion.
\subsection{Relation to Other Sparse Recovery Algorithms}

The proposed iterative RRMMSE estimator shares its greedy support-growing structure with three families of algorithms in the sparse recovery literature.
\paragraph*{Matching Pursuit and OMP}
MP and OMP~\cite{Tropp_Gilbert_2007} select at each iteration the column of the sensing matrix that has the largest inner product with the current residual, and either project the residual onto the chosen column (MP) or onto the span of all chosen columns(OMP).
These algorithms operate on a deterministic least-squares criterion and do not carry a posterior covariance, so they cannot update a Bayesian prior or provide a Bayesian stopping criterion analogous to $\tfrac{1}{M}\tr(\Ke^{(i)})$.
\paragraph*{CLEAN and Active CLEAN}
The CLEAN algorithm~\cite{Bose_2012} subtracts a scaled point-spread response of the brightest scatterer from the residual and iterates.
CLEAN is most effective when the point-spread function is well concentrated;
for non-contiguous transmission spectra the point-spread function has high integrated sidelobes, and CLEAN performance degrades with increasing spectral sparsity.
Active CLEAN~\cite{Bose_2012} partially addresses this by adapting the subtracted response, but still operates on a deterministic least-squares criterion.
\paragraph*{Sparse Bayesian Learning}
Earlier SBL work ~\cite{Wipf_Rao_2004,Tipping_2001} utilized independent Gaussian priors on the $\gamma_m$ with separately learned variances and estimated these variances by evidence maximization.
The proposed estimator can be viewed as a single-pass online version of SBL in which the prior variance $[\Kg]_{m,m}$ is updated to the closed-form expression in~\eqref{eq:Kg_update} after each new
range bin is added to the support, instead of being re-estimated by a full evidence maximization at every iteration.
This single-pass structure trades a small amount of estimation accuracy for a fixed per-iteration cost.
The principal differences between the proposed algorithm and the methods above are: (i)~the use of the Bayesian MMSE filter at each iteration, which preserves the posterior error covariance $\Ke^{(i)}$ and provides a natural stopping criterion;
(ii)~the prior update~\eqref{eq:Kg_update} which gives the estimator its reduced-rank behavior and which differs from OMP and MP in that previously selected reflection coefficients retain a posterior variance rather than being treated as exact;
and (iii)~compatibility with the MFI sparse spectrum design of~\cite{Tan_2016,Stiles_2009}, which guarantees that the sparse sensing matrix has a coarray with few holes and few redundancies.
\section{Simulation Results and Discussion}\label{sec:results}
\subsection{Simulation Setup}

A pulsed radar signal is simulated with bandwidth $B = 20$~kHz, delay-spectrum timewidth $T_0 = 10$~ms, and Nyquist oversampling factor of $10$, yielding $M = 2 B T_0 + 1 = 401$ range bins and $M_{\mathrm{nyq}} = 4000$ candidate frequency samples on the oversampled grid.
The MFI spectrum design uses a block size of $1.25\%$ of the Nyquist count, producing $50\%$ and $75\%$ spectrally occupied supports as shown in Fig.~\ref{fig:50_spectrum} and~\ref{fig:75_spectrum}.
Detectable scatterers are placed randomly in the range profile to occupy a fraction $\rho \in \{10\%,20\%,30\%,40\%,50\%,60\%,100\%\}$ of the $M-1$ resolution cells.
The magnitudes of the reflection coefficients are drawn uniformly on a logarithmic scale spanning $[-10,30]$~dB and the phases are drawn uniformly on $[0,2\pi)$.
The noise variance is set such that the per-measurement SNR is constant across sparsity levels.
The a~priori target variance is $\sigma_\gamma^2 = 5\times 10^3$ and the stopping tolerance is set so that the algorithm terminates when $\tfrac{1}{M}\tr(\Ke^{(i)})$ ceases to decrease.
\subsection{Matched-Filter Baseline}

\begin{figure}[!t]
\centerline{\includegraphics[width=3.4in]{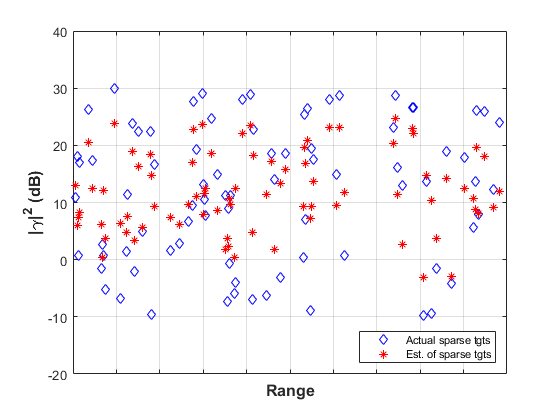}}
\caption{Matched-filter estimate of the range profile for $20\%$ target occupancy and $50\%$ spectrally occupied MFI-designed spectrum.
Green stars: matched-filter estimate at all $M$ range bins. Blue diamonds: true $|\gamma_m|^2$ at the target locations.
Red stars: matched-filter estimate at the target locations.}
\label{fig:mf_20_50}
\end{figure}

\begin{figure}[!t]
\centerline{\includegraphics[width=3.4in]{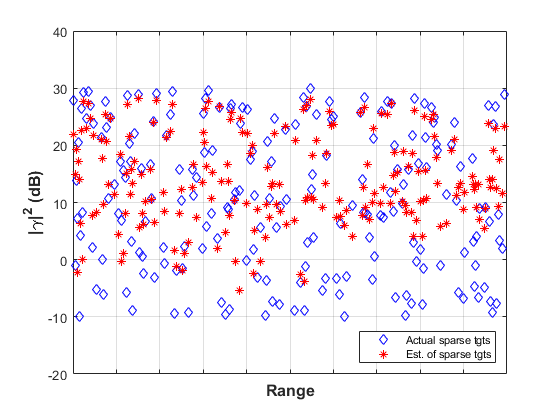}}
\caption{Matched-filter estimate of the range profile for
$50\%$ target occupancy and $75\%$ spectrally occupied
MFI-designed spectrum.
Marker convention as in
Fig.~\ref{fig:mf_20_50}.}
\label{fig:mf_50_75}
\end{figure}

Figs.~\ref{fig:mf_20_50} and~\ref{fig:mf_50_75} show the matched-filter estimate $\hat\gam_{\mathrm{MF}} = \HH\herm\vv$ of the range profile, compared with the true profile, for ($\rho{=}20\%$, $50\%$ occupancy) and ($\rho{=}50\%$, $75\%$ occupancy).
The high integrated sidelobe level produced by the non-contiguous spectrum prevents matched filtering from recovering the support of $\gam$.
Weak scatterers are masked entirely by sidelobe leakage from stronger scatterers, even at $75\%$ spectral occupancy.

\subsection{Iteration Snapshots and Final Estimate}

\begin{figure}[!t]
\centerline{\includegraphics[width=3.4in]{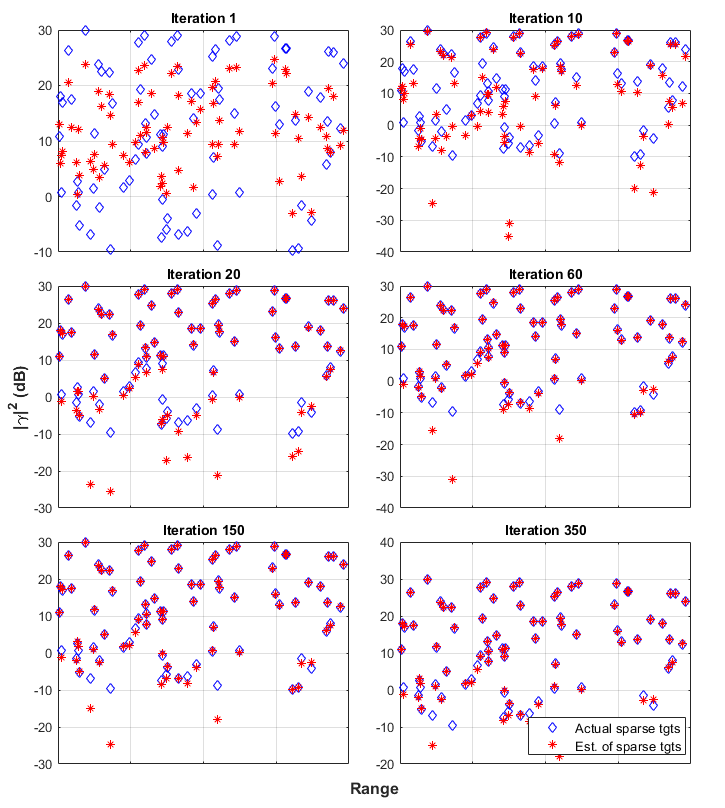}}
\caption{Snapshots of the iterative RRMMSE estimate of the range
profile at three iterations for $20\%$ target occupancy and
$50\%$ spectrally occupied MFI-designed spectrum.
Marker
convention as in Fig.~\ref{fig:mf_20_50}.}
\label{fig:snapshots}
\end{figure}

Fig.~\ref{fig:snapshots} shows three snapshots of the iterative
estimate $\hat\gam^{(i)}$ at iterations $i=1$, $i=G/2$, and $i=G$
for the case ($\rho{=}20\%$, $50\%$ occupancy).
The first snapshot
is the matched-filter initialization; the middle snapshot shows
the support being progressively identified;
the last snapshot
shows the converged estimate after the algorithm has absorbed the
significant scatterers into $\Kg$.

\begin{figure}[!t]
\centerline{\includegraphics[width=3.4in]{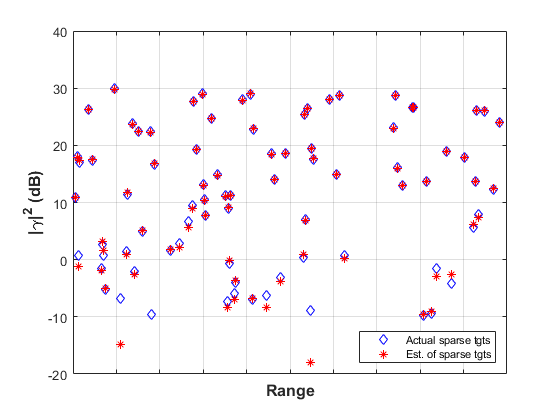}}
\caption{Final iterative RRMMSE estimate of the range profile
after convergence for $20\%$ target occupancy and $50\%$
spectrally occupied MFI-designed spectrum.
The algorithm recovers
nearly all significant scatterers with error close to the noise
floor.}
\label{fig:rrmmse_20_50}
\end{figure}

\begin{figure}[!t]
\centerline{\includegraphics[width=3.4in]{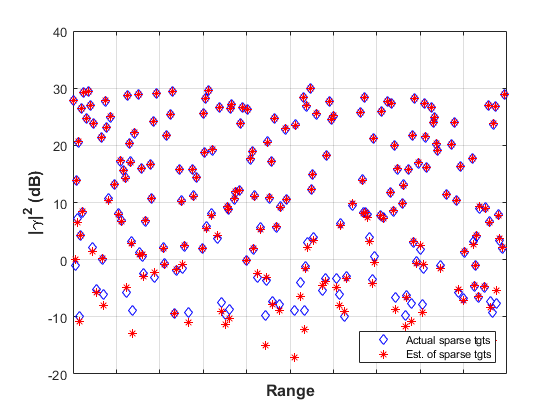}}
\caption{Final iterative RRMMSE estimate of the range profile
after convergence for $50\%$ target occupancy and $75\%$
spectrally occupied MFI-designed spectrum.
Only the smallest
scatterers (below the noise floor of the $75\%$ spectrum) are
missed.}
\label{fig:rrmmse_50_75}
\end{figure}

Figs.~\ref{fig:rrmmse_20_50} and~\ref{fig:rrmmse_50_75} show the
final iterative RRMMSE estimate after convergence for the same
two scenarios as in Figs.~\ref{fig:mf_20_50}
and~\ref{fig:mf_50_75}.
The iterative algorithm recovers
essentially all significant scatterers in both scenarios.
The few
errors that remain occur for the smallest reflection coefficients,
which sit below the noise floor imposed by the residual integrated
sidelobe level of the sparse spectrum.
\subsection{Convergence of MSE}

\begin{figure}[!t]
\centerline{\includegraphics[width=3.4in]{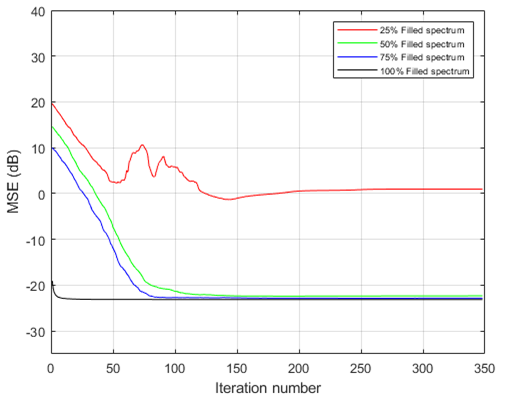}}
\caption{MSE of the range-profile estimate versus iteration index
for $20\%$ target occupancy and varying spectral occupancy
($50\%$, $75\%$, $100\%$).
Solid curves: $\mathrm{MSE}_{\Ke}$
computed from the trace of $\Ke^{(i)}$. Dashed curves:
$\mathrm{MSE}_{\mathrm{GT}}$ computed against ground truth. All
three configurations converge to nearly identical final MSE,
confirming that the MFI-designed sparse spectra retain almost the
full estimation accuracy of the contiguous baseline when the
target profile is sparse.}
\label{fig:mse_vs_iter_targets}
\end{figure}

\begin{figure}[!t]
\centerline{\includegraphics[width=3.4in]{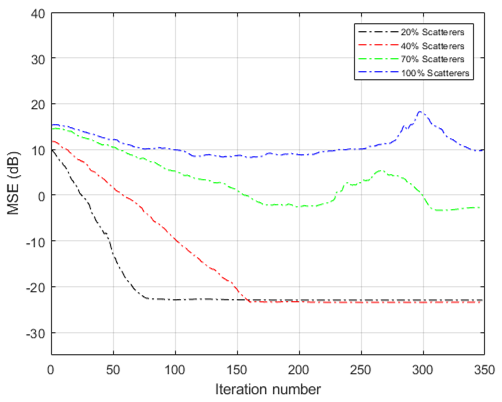}}
\caption{MSE of the range-profile estimate versus iteration index
for $75\%$ spectrally occupied MFI-designed spectrum and varying
target occupancy. The $100\%$ target-occupancy curve does not
converge because the support of $\gam$ exceeds the rank $K$ of
the sparse sensing matrix.}
\label{fig:mse_vs_iter_spectrum}
\end{figure}

The first MSE is evaluated from the posterior error covariance,
\begin{equation}
\mathrm{MSE}_{\Ke}^{(i)}
\;=\;
\frac{1}{M}\,\tr\!\left(\Ke^{(i)}\right),
\label{eq:mse_Ke}
\end{equation}
which, by the connection in~\eqref{eq:mfi_in_Ke}, is also the
running Bayesian CRLB on $\gam$ achieved by the spectrum after
$i$~iterations of the support-growing procedure.
The second MSE is evaluated against the ground truth $\gam$,
\begin{equation}
\mathrm{MSE}_{\mathrm{GT}}^{(i)}
\;=\; \frac{1}{M}\,(\gam-\hat\gam^{(i)})\herm(\gam-\hat\gam^{(i)}).
\label{eq:mse_GT}
\end{equation}
Fig.~\ref{fig:mse_vs_iter_targets} shows both quantities as a function of iteration for fixed target occupancy $\rho=20\%$ and three different spectral occupancies: $50\%$, $75\%$, and $100\%$~(fully filled).
Fig.~\ref{fig:mse_vs_iter_spectrum} shows the same quantities for fixed $75\%$ spectral occupancy and target occupancies ranging from $10\%$ to $100\%$.

When the number of detectable scatterers $G$ satisfies $G \leq K$, the algorithm converges in $G$ iterations and the converged MSE is within a small fraction of a dB of the contiguous-spectrum baseline. 
As $\rho$ approaches the spectral occupancy, the effective rank saturates and the per-iteration reduction in posterior covariance becomes progressively smaller, which is reflected in the flattening of the convergence curves of Fig.~\ref{fig:mse_vs_iter_spectrum} prior to termination. 
When $\rho$ exceeds the spectral occupancy (the $100\%$-target curve in Fig.~\ref{fig:mse_vs_iter_spectrum}), the system is fundamentally underdetermined and the algorithm fails to converge, consistent with the rank constraint $G\leq K$ noted in Section~\ref{sec:formulation}.

\subsection{Summary}

The simulation results establish that an MFI-designed sparse spectrum at $50\%$ occupancy is sufficient to recover a sparse range profile with up to $40\%$ significant scatterers at MSE close to the contiguous-spectrum baseline. 
A half of the radar bandwidth can therefore be relinquished to other wireless users with negligible loss of estimation accuracy, provided the radar scene is itself sparse. 
The iterative RRMMSE algorithm provides a tractable implementation path: the per-iteration cost is dominated by a single $K\times K$ inversion, and the algorithm terminates naturally when the posterior covariance ceases to decrease.
The accurate recovery of the complex reflection coefficient ($\gamma_m$) has implications beyond basic target detection. 
In quantitative remote sensing, the precise amplitude of the recovered target~\cite{Talasila_RADAR_2025} is a prerequisite for radiometric calibration and geophysical parameter inversion~\cite{Talasila_PhD_2025}, and for the glaciological products derived downstream from calibrated echograms, such as subglacial-landscape and ice-dynamics reconstructions~\cite{Hoffman_NatGeo_2023}. 
The ability of the RRMMSE algorithm to preserve the amplitude fidelity of significant scatterers near the contiguous-spectrum baseline indicates that sparse transmission architectures can support rigorously calibrated radar data products.

\section{Conclusion}\label{sec:conclusion}

An iterative reduced-rank MMSE algorithm has been proposed for the recovery of sparse range profiles from non-contiguous radar transmission spectra designed by the MFI criterion. 
The Bayesian CRLB on the per-bin reflection coefficient has been derived for the non-contiguous spectrum model, and its trace has been shown to be the integrand of the MFI design criterion, which ties the receive-side estimator directly to the spectrum-design objective. 
The computational complexity of the proposed estimator has been shown to scale as $\Order(G^2 M K^2)$, which is substantially cheaper than the $\Order(M^3)$ cost of a one-step full-rank MMSE when the radar scene is sparse, and the same recursion delivers a posterior error covariance that doubles as a stopping criterion for the iteration. 
Simulations on MFI-designed spectra at $50\%$ and $75\%$ occupancy show that the algorithm recovers sparse range profiles with MSE close to the contiguous-spectrum baseline, provided the number of significant scatterers does not exceed the rank of the sparse sensing matrix. 
These results indicate that up to half of the radar bandwidth can be relinquished to other wireless users with negligible loss of estimation accuracy whenever the radar scene admits a sparse representation. 
This is a useful operating margin for spectrum-constrained airborne and ground-based radar systems, and it provides a quantitative basis on which a cognitive transmitter may negotiate spectrum access without committing to a fixed allocation in advance. 

In future, the framework can be extended to a two-dimensional support estimation by treating direction-of-arrival jointly with range, and then to three-dimensional range-Doppler-DoA estimation by applying the same MFI-driven support growth across coupled spectral, angular, and slow-time sampling grids. 
A natural application of such a multidimensional extension is the multipass estimation of ice-sheet vertical velocity, englacial layer geometry, and englacial deformation, where the support of significant scatterers must be tracked jointly across range, angle, and repeated passes~\cite{Ariho_PhD_2023,Ariho_PAST2022,Holschuh_AGU_2022}.
Future work could include a hardware-in-the-loop verification on a software-defined radar testbed instrumented with a real-time MFI optimizer in the transmit chain and the iterative RRMMSE estimator in the receive chain, closing the cognitive feedback loop between waveform generation and range-profile estimation.

\section*{Acknowledgment}

The authors acknowledge support from the Center for Surveillance Research, and the Radar Systems and Remote Sensing Laboratory at the University of Kansas.


\section*{Appendix}

\begin{table}[h]
\centering
\caption{List of Symbols and Variables}
\label{tab:symbols}
\begin{tabular}{@{}ll@{}}
\toprule
\textbf{Symbol} & \textbf{Description} \\ \midrule
$B$ & Radar bandwidth \\
$f_c$ & Center frequency \\
$N$ & Total number of uniformly spaced frequencies in the nominal bandwidth \\
$\Delta\omega$ & Angular frequency spacing \\
$S(\omega)$ & Transmit spectrum \\
$s_n$ & Complex weight of the $n$-th spectral line \\
$K$ & Number of preserved (nonzero) spectral lines \\
$M$ & Number of range resolution cells \\
$\tau_m$ & Center delay of the $m$-th range cell \\
$\gamma_m$ & Complex reflection coefficient of the $m$-th range cell \\
$\mathbf{v}$ & Discrete radar measurement vector ($K \times 1$) \\
$\mathbf{H}$ & Sensing matrix ($K \times M$) \\
$\boldsymbol{\gamma}$ & Range-profile vector ($M \times 1$) \\
$\mathbf{n}$ & Additive measurement noise vector ($K \times 1$) \\
$\omega_k$ & Preserved angular frequencies \\
$\mathbf{K}_n$ & Noise covariance matrix ($\sigma_n^2 \mathbf{I}_K$) \\
$\sigma_n^2$ & Noise variance / noise power \\
$\mathbf{I}_K, \mathbf{I}_M$ & Identity matrices of size $K$ and $M$ \\
$M_{\mathrm{nyq}}$ & Number of Nyquist frequency samples in $B$ \\
$G$ & Number of resolution cells with significant reflectivity \\
$\mathbf{K}_\gamma$ & A priori target covariance matrix ($\sigma_\gamma^2 \mathbf{I}_M$) \\
$\sigma_\gamma^2$ & A priori target variance \\
$\mathbf{W}_{\mathrm{MMSE}}$ & Minimum Mean-Square Error estimator weight matrix \\
$\hat{\boldsymbol{\gamma}}_{\mathrm{MMSE}}$ & MMSE estimate of the range-profile vector \\
$\mathbf{K}_\epsilon$ & Posterior error covariance matrix \\
$\mathbf{J}(\boldsymbol{\gamma})$ & Deterministic Fisher Information Matrix \\
$\mathbf{J}_B(\boldsymbol{\gamma})$ & Bayesian Fisher Information Matrix \\
$\Delta\mathbf{I}(K)$ & Marginal Fisher Information matrix \\
$\mathrm{MFI}(K)$ & Scalar Marginal Fisher Information criterion \\
$c(l)$ & Co-array function at lag $l$ \\
$w_n$ & Binary indicator encoding presence/absence of $n$-th spectral line \\
$i$ & Iteration index \\
$\mathcal{S}^{(i)}$ & Estimated target support set at iteration $i$ \\
$\mathcal{I}_m$ & Augmented index set containing candidate bin $m$ \\
$\eta$ & Stopping tolerance for iteration algorithm \\
$G_{\max}$ & Maximum iterations \\
$\rho$ & Target occupancy fraction \\
$T_0$ & Delay-spectrum time \\
\bottomrule
\end{tabular}
\end{table}


\begin{thebibliography}{99}

\bibitem{Griffiths_2015} H.~Griffiths, L.~Cohen, S.~Watts, E.~Mokole, C.~Baker, M.~Wicks, and S.~Blunt, ``Radar spectrum engineering and management: Technical and regulatory issues,'' \emph{Proc. IEEE}, vol.~103, no.~1, pp.~85--102, Jan. 2015.

\bibitem{Haykin_2006} S.~Haykin, ``Cognitive radar: A way of the future,'' \emph{IEEE Signal Process. Mag.}, vol.~23, no.~1, pp.~30--40, Jan. 2006.

\bibitem{FCC_3p5GHz_2015} Federal Communications Commission, ``Amendment of the Commission's rules with regard to commercial operations in the 3550--3650~MHz band,'' GN Docket No.~12-354, Rep. and Order and Second Further NPRM, FCC 15-47, Apr. 2015.

\bibitem{Cohen_2018} D.~Cohen, K.~V. Mishra, and Y.~C. Eldar, ``Spectrum sharing radar: Coexistence via Xampling,'' \emph{IEEE Trans. Aerosp. Electron. Syst.}, vol.~54, no.~3, pp.~1279--1296, Jun. 2018.

\bibitem{Chintareddy_GLOBECOM_2023} S.~R. Chintareddy, K.~Roach, K.~Cheung, and M.~Hashemi, ``Collaborative Wideband Spectrum Sensing and Scheduling for Networked UAVs in UTM Systems,'' in \emph{Proc. IEEE Global Commun. Conf. (GLOBECOM)}, Kuala Lumpur, Malaysia, Dec. 2023, pp.~3064--3069.

\bibitem{Chintareddy_TMLCN_2025} S.~R. Chintareddy, K.~Roach, K.~Cheung, and M.~Hashemi, ``Federated Learning-Based Collaborative Wideband Spectrum Sensing and Scheduling for UAVs in UTM Systems,'' \emph{IEEE Trans. Mach. Learn. Commun. Netw.}, vol.~3, pp.~296--314, Feb. 2025, doi:~10.1109/TMLCN.2025.3540747.

\bibitem{Morales_IGARSS_2024_Thwaites} F.~R. Morales \emph{et al.}, ``An UWB UHF Ice-penetrating Radar for the Thwaites MELT Project,'' in \emph{Proc. IEEE Int. Geosci. Remote Sens. Symp. (IGARSS)}, Athens, Greece, Jul. 2024.

\bibitem{Morales_IGARSS_2024_Multichannel} F.~R. Morales \emph{et al.}, ``A Multi-channel UWB Airborne Radar for Swath Mapping of Snow Layers,'' in \emph{Proc. IEEE Int. Geosci. Remote Sens. Symp. (IGARSS)}, Athens, Greece, Jul. 2024.

\bibitem{Talasila_IGARSS_2023} H.~M. Talasila \emph{et al.}, ``High-Altitude Measurements of Snow Thickness using Ultra-Wideband Microwave Radar,'' in \emph{Proc. IEEE Int. Geosci. Remote Sens. Symp. (IGARSS)}, Pasadena, CA, USA, Jul. 2023, pp.~24--27.

\bibitem{Talasila_RadarConf_2024} H.~M. Talasila and J.~Paden, ``Crossover Analysis for the Radiometric Calibration of Radar Depth Sounder Data Products,'' in \emph{Proc. IEEE Radar Conf. (RadarConf)}, Denver, CO, USA, May 2024.

\bibitem{Aubry_2014} A.~Aubry, A.~De~Maio, M.~Piezzo, and A.~Farina, ``Radar waveform design in a spectrally crowded environment via nonconvex quadratic optimization,'' \emph{IEEE Trans. Aerosp. Electron. Syst.}, vol.~50, no.~2, pp.~1138--1152, Apr. 2014.

\bibitem{Blunt_Mokole_2016} S.~D. Blunt and E.~L. Mokole, ``Overview of radar waveform diversity,'' \emph{IEEE Aerosp. Electron. Syst. Mag.}, vol.~31, no.~11, pp.~2--42, Nov. 2016.

\bibitem{Liu_2020} F.~Liu, C.~Masouros, A.~P. Petropulu, H.~Griffiths, and L.~Hanzo, ``Joint radar and communication design: Applications, state-of-the-art, and the road ahead,'' \emph{IEEE Trans. Commun.}, vol.~68, no.~6, pp.~3834--3862, Jun. 2020.

\bibitem{Liu_2022} F.~Liu, Y.~Cui, C.~Masouros, J.~Xu, T.~X. Han, Y.~C. Eldar, and S.~Buzzi, ``Integrated sensing and communications: Toward dual-functional wireless networks for 6G and beyond,'' \emph{IEEE J. Sel. Areas Commun.}, vol.~40, no.~6, pp.~1728--1767, Jun. 2022.

\bibitem{Mishra_2019} K.~V. Mishra, M.~R. Bhavani~Shankar, V.~Koivunen, B.~Ottersten, and S.~A. Vorobyov, ``Toward millimeter-wave joint radar communications: A signal processing perspective,'' \emph{IEEE Signal Process. Mag.}, vol.~36, no.~5, pp.~100--114, Sep. 2019.

\bibitem{Candes_2006} E.~J. Cand\`es, J.~Romberg, and T.~Tao, ``Robust uncertainty principles: Exact signal reconstruction from highly incomplete frequency information,'' \emph{IEEE Trans. Inf. Theory}, vol.~52, no.~2, pp.~489--509, Feb. 2006.

\bibitem{Ender_2010} J.~H.~G. Ender, ``On compressive sensing applied to radar,'' \emph{Signal Process.}, vol.~90, no.~5, pp.~1402--1414, May 2010.

\bibitem{Baraniuk_2007} R.~G. Baraniuk, ``Compressive sensing,'' \emph{IEEE Signal Process. Mag.}, vol.~24, no.~4, pp.~118--121, Jul. 2007.

\bibitem{Quan_2024} Y.~Quan, Y.~Wu, L.~Duan, G.~Xu, M.~Xue, Z.~Liu, and M.~Xing, ``A review of radar signal processing based on sparse recovery,'' \emph{J. Radars}, vol.~13, no.~1, pp.~46--67, Feb. 2024.

\bibitem{Greco_2018} M.~S. Greco, F.~Gini, P.~Stinco, and K.~Bell, ``Cognitive radars: On the road to reality: Progress thus far and possibilities for the future,'' \emph{IEEE Signal Process. Mag.}, vol.~35, no.~4, pp.~112--125, Jul. 2018.

\bibitem{Charlish_Hoffmann_2020} A.~Charlish and F.~Hoffmann, ``Cognitive radar management,'' in \emph{Novel Radar Techniques and Applications}, R.~Klemm, U.~Nickel, C.~Gierull, P.~Lombardo, H.~Griffiths, and W.~Koch, Eds. Stevenage, U.K.: SciTech Publishing, 2017, ch.~5, pp.~157--193.

\bibitem{Ariho_PhD_2023} G.~Ariho, ``Multipass SAR Processing for Ice Sheet Vertical Velocity and Tomography Measurements,'' Ph.D. dissertation, Dept. Elect. Eng. Comput. Sci., University of Kansas, Lawrence, KS, USA, 2023.

\bibitem{Ariho_PAST2022} G.~Ariho, J.~Paden, A.~Hoffman, K.~Christianson, and N.~D. Holschuh, ``Joint estimation of ice sheet vertical velocity and englacial layer geometry from multipass synthetic aperture radar data,'' in \emph{Proc. IEEE Int. Symp. Phased Array Systems and Technology (PAST)}, 2022.

\bibitem{Ariho_IGARSS2020} B.~Miller, G.~Ariho, J.~Paden, and E.~Arnold, ``Multipass SAR Processing for Radar Depth Sounder Clutter Suppression, Tomographic Processing, and Displacement Measurements,'' in \emph{Proc. IEEE Int. Geosci. Remote Sens. Symp. (IGARSS)}, 2020, pp.~822--825.

\bibitem{Holschuh_AGU_2022} N.~Holschuh, G.~Ariho, K.~A. Christianson, A.~O. Hoffman, and J.~D. Paden, ``Catchment-Scale Measurement of Englacial Deformation in Greenland,'' in \emph{AGU Fall Meeting Abstracts}, 2022, abstract C55B-0407.

\bibitem{Tan_2016} P.~S. Tan, J.~M. Stiles, and S.~D. Blunt, ``Optimizing sparse allocation for radar spectrum sharing,'' in \emph{Proc. IEEE Radar Conf. (RadarConf)}, Philadelphia, PA, USA, May 2016, pp.~1--6.

\bibitem{Stiles_2009} J.~M. Stiles and J.~Jenshak, ``Sparse array construction using marginal Fisher's information,'' in \emph{Proc. Int. Waveform Diversity and Design Conf.}, Kissimmee, FL, USA, Feb. 2009, pp.~119--123.

\bibitem{Holm_1989} S.~Holm and J.-F. Hopperstad, ``The coarray of sparse arrays with minimum sidelobe level,'' in \emph{Proc. IEEE Nordic Signal Process. Symp. (NORSIG)}, 1998, pp.~137--140.

\bibitem{Moffet_1968} A.~T. Moffet, ``Minimum-redundancy linear arrays,'' \emph{IEEE Trans. Antennas Propag.}, vol.~16, no.~2, pp.~172--175, Mar. 1968.

\bibitem{Paden_OPR_2022} J.~Paden \emph{et al.}, ``Open Polar Radar: Standardizing Radargram Referencing, Processing and Access,'' in \emph{First Int. Workshop on Antarctic RINGS}, Troms\o{}, Norway, Jun. 2022.

\bibitem{Paden_AGU_2021} J.~Paden \emph{et al.}, ``Open Polar Radar Software and Services to Standardize Radar Echograms and Integrate into a Geospatial Database,'' in \emph{AGU Fall Meeting Abstracts}, vol.~2021, pp.~C51A-08, Dec. 2021.

\bibitem{Oluwanisola_RadarConf_2023} I.~Oluwanisola \emph{et al.}, ``Snow Radar Echogram Layer Tracker: Deep Neural Networks for Radar Data from NASA Operation IceBridge,'' in \emph{Proc. IEEE Radar Conf. (RadarConf)}, San Antonio, TX, USA, May 2023.

\bibitem{Oluwanisola_SRED_2025} I.~Oluwanisola \emph{et al.}, ``AI-ready Snow Radar Echogram Dataset (SRED) for climate change monitoring,'' \emph{arXiv preprint arXiv:2505.00786}, pp.~1--12, May 2025.

\bibitem{Rowell_Climate_2023} I.~Rowell \emph{et al.}, ``An age scale for new climate records from Sherman Island, West Antarctica,'' \emph{Climate of the Past}, vol.~19, no.~8, pp.~1699--1714, Aug. 2023.

\bibitem{Hoffman_NatGeo_2023} A.~O. Hoffman, N.~Holschuh, M.~Mueller, J.~Paden, A.~Muto, G.~Ariho, C.~Brigham, J.~E. Christian, L.~Davidge, E.~Heitmann, B.~Hills, A.~Horlings, S.~Morey, G.~O'Connor, T.~J. Fudge, E.~J. Steig, and K.~Christianson, ``Scars of tectonism promote ice-sheet nucleation from Hercules Dome into West Antarctica,'' \emph{Nature Geoscience}, vol.~16, no.~11, pp.~1005--1013, Nov. 2023.

\bibitem{Tropp_Gilbert_2007} J.~A. Tropp and A.~C. Gilbert, ``Signal recovery from random measurements via orthogonal matching pursuit,'' \emph{IEEE Trans. Inf. Theory}, vol.~53, no.~12, pp.~4655--4666, Dec. 2007.

\bibitem{Bose_2012} R.~Bose, ``Active CLEAN: A modified CLEAN algorithm for HRRPs of contiguous targets with thinned spectrum,'' \emph{IEEE Trans. Aerosp. Electron. Syst.}, vol.~48, no.~2, pp.~930--939, Apr. 2012.

\bibitem{Wipf_Rao_2004} D.~P. Wipf and B.~D. Rao, ``Sparse Bayesian learning for basis selection,'' \emph{IEEE Trans. Signal Process.}, vol.~52, no.~8, pp.~2153--2164, Aug. 2004.

\bibitem{Tipping_2001} M.~E. Tipping, ``Sparse Bayesian learning and the relevance vector machine,'' \emph{J. Mach. Learn. Res.}, vol.~1, pp.~211--244, Jun. 2001.

\bibitem{Talasila_RADAR_2025} H.~M. Talasila and J.~Paden, ``Natural Target Analysis for the Radiometric Calibration of Radar Depth Sounder Data Products,'' in \emph{Proc. IEEE Int. Radar Conf. (RADAR)}, Atlanta, GA, USA, May 2025.

\bibitem{Kay_1993} S.~M. Kay, \emph{Fundamentals of Statistical Signal Processing, Vol.~I: Estimation Theory}. Englewood Cliffs, NJ, USA: Prentice-Hall, 1993.

\bibitem{Talasila_PhD_2025} H.~M. Talasila, ``Radiometric Calibration of Radar Depth Sounder Data Products,'' Ph.D. dissertation, Dept. Elect. Eng. Comp. Sci., Univ. Kansas, Lawrence, KS, USA, 2025.

\end{thebibliography}
\end{document}